\documentclass[preprint]{aastex}

\begin{document}

\title{Orbits and photometry of Pluto's satellites: Charon, S/2005 P1 and S/2005 P2}

\shorttitle{Pluto satellites}

\author{Marc W. Buie, William M. Grundy}

\affil{Lowell Observatory, 1400 W. Mars Hill Rd., Flagstaff, AZ 86001}
\email{buie@lowell.edu, grundy@lowell.edu}

\author{Eliot F. Young, Leslie A. Young, and S. Alan Stern}
\affil{Southwest Research Institute, Boulder, CO}
\email{efy@boulder.swri.edu, layoung@boulder.swri.edu, alan@boulder.swri.edu}

\slugcomment{Version 1.0, Submitted 2005/12/19}

\begin{abstract}

We present new astrometry of Pluto's three satellites from images
taken of the Pluto system during 2002-3 with the High Resolution Camera
(HRC) mode of the Advanced Camera for Surveys (ACS) instrument on the
Hubble Space Telescope.  The observations were designed to produce
an albedo map of Pluto but they also contain images of Charon and
the two recently discovered satellites, S/2005 P1 and S/2005 P2\null.
Orbits fitted to all three satellites are co-planar and, for Charon and P2,
have eccentricities consistent with zero.  The orbit of the outermost
satellite, P1, has a significant eccentricity of 0.0052 $\pm$ 0.0011.
Orbital periods of P1, P2, and Charon are 38.2065 $\pm$ 0.0014, 24.8562
$\pm$ 00013, and 6.3872304 $\pm$ 0.0000011 days, respectively.  The total
system mass based on Charon's orbit is $1.4570 \pm 0.0009$~x~$10^{22}$~kg.
We confirm previous results that orbital periods are close to the ratio
of 6:4:1 (P1:P2:Charon) indiciative of mean-motion resonances, but
our results formally preclude precise integer period ratios.  The orbits
of P1 and P2, being about the barycenter rather than Pluto, enable us
to measure the Charon/Pluto mass ratio as 0.1165$\pm$0.0055.  This new
mass ratio implies a density of 1.66 $\pm$ 0.06 g~cm$^{-3}$ for Charon
and 2.03 $\pm$ 0.06 g~cm$^{-3}$ for Pluto thus adding confirmation that
Charon is somewhat under-dense relative to Pluto.  Finally, by stacking
all images, we can extract globally averaged photometry.  P1 has a mean
opposition magnitude of $V=24.39 \pm 0.02$ and color of $(B-V) = 0.644
\pm 0.028$.  P2 has a mean opposition magnitude of $V=23.38 \pm 0.02$
and color of $(B-V) = 0.907 \pm 0.031$.  The colors indicate that P1 is
spectrally neutral and P2 is slightly more red than Pluto.  The variation
in surface color with radial distance from Pluto is quite striking (red,
neutral, red, neutral) and begs further study.

\end{abstract}



\def\capellipses{		
Sky plane observations (points with error bars) and predicted positions
(open circles) for P1 and P2, based on our best fit orbits.  Observations
of Charon are also shown, but the predicted positions are omitted to
avoid clutter.  Gray ellipses show the instantaneous orbits about the
barycenter at a single, arbitrary epoch for Pluto, Charon, P2, and P1,
from smallest to largest, respectively.  The \citet{wea05} measurements
are distinguished by double circles for their predicted positions.}

\def\cappoles{			
One sigma contours of orbit poles on the J2000 sky plane for our best fit
orbits for Charon, P1, and P2, compared with the \citet{tho97} orbit pole
for Charon (``TB97'')\null.  These contours were computed from a large set
of orbits where each orbital element is drawn from a gaussian distribution
consistent with the element and its associated uncertainties.}

\def\capresiduals{		
East-West and North-South residuals relative to our best fit orbits
plotted versus orbital longitude for Charon (upper two panels, with open
circles representing the 60 data points from \citet{tho97} and diamonds
representing our 394 new observations) and for P1 and P2 (represented
by asterisks and diamonds, respectively, in the lower two panels, with
\citet{wea05} data points circled).}

\def\capchisq{			
Mass ratio slices in reduced $\chi^2$ space for P1 and P2 (dashed
and dotted curves, respectively), showing 1 sigma contours, defined as
$\chi^2_{\rm minimum}+1$ \citep{numrec92}.  The solid curve is for P1 and
P2 combined.  The \citet{olk03} mass ratio measurement is also indicated.
The fact that reduced $\chi^2$ levels are somewhat below unity suggests
that our adopted astrometric uncertainties ($\pm$0.009~arcsec for P1
and $\pm$0.015~arcsec for P2) are slightly too conservative.}

\def\capstack{			
Final stacked images for S/2005 P1 and P2.  The left panels show the
stacked images for the F435W filter.  The right panels show the F555W
filter data.  The pair of images on the top are stacked based on the
our fitted ephemeris motion for S/2005 P1 while the pair on the bottom
were stacked on the S/2005 P2 ephemeris.  The per-pixel sky noise is
0.3 e-/sec and is completely read-noise limited.  The peak signals on
the satelliates in each image are P1:F435W=2.93~e-/sec,
P1:F555W=2.22~e-/sec, P2:F435W=1.89~e-/sec, and P2:F555W=2.19~e-/sec.
The display stretch is the same for all four stacked images and is set
to $-3$ to $+10 \sigma$ of the sky level.}

\section{Introduction}

The study of Pluto was greatly facilitated in 1978 with the discovery
of its first satellite, Charon \citep{chr78}\null.  That discovery made
possible the accurate determination of its mass which had previously
been largely a matter of conjecture.  Later, in the late 1980's, Pluto
studies were transformed by the mutual events between Pluto and Charon
\citep[e.g.,][]{bui92, bin97, you01}\null.  Charon remains an interesting
object it is own right, but its role as a tool from which to understand
the system should not be understated.

Two new moons were recently discovered in orbit around Pluto
\citep{wea05}\null.  More precisely, they orbit the center of mass
of the system, which is very close to the Pluto-Charon barycenter.
As with Charon, these new objects will be studied in their own right
and will also be useful as probes or test masses in the Pluto system.
Given astrometry of sufficient precision and time-base, one can now
easily deduce the precise Charon\slash Pluto mass ratio.  One might also
hope to determine the masses of the new satellites through their mutual
perturbations.  However, their mutual gravitational force is more than
3 orders of magnitude weaker than the force exerted on them by Pluto
and Charon, so the dynamics of their presumably resonant orbits may well
completely mask any measurable effect P1 and P2 may have on each other.

The preliminary orbits computed by \citet{wea05} were based on just two
epochs of data separated by only three days, much less than a full orbit
of either satellite.  Also, the data were derived from images where Pluto
and Charon were both saturated.  These constraints led to a restricted
solution for the orbit where it was assumed that the objects were in
circular orbits in the same orbital plane as Charon.  As we shall show,
this assumption turned out to be very close to the correct answer.

The data presented in this work are derived from pre-discovery HST
observations that span multiple orbits of all satellites and do so with
images where Pluto and Charon are not saturated.  This paper presents
the first unrestricted fits to the orbits of the new satellites using
pre-discovery HST observations.

\section{Observations}

Images were taken of the Pluto system from June 2002 to June 2003 with the
High Resolution Camera (HRC) mode of the Advanced Camera for Surveys (ACS)
on the Hubble Space Telescope.  The observations were designed to permit
construction of a photometrically accurate map of the surface of Pluto.
A total of 12 visits were allocated and scheduled to occur at specific
sub-earth longitudes of Pluto at a 30-degree rotational resolution.
The geometric circumstances of the observations are tabulated in
Table~\ref{table_circumstances}\null.  Each visit was designed to fit in
a single visibility window but scheduling constraints stretched the time
line out beyond a single orbit.  Within each visit, two filters were used.
F435W and F555W filters were chosen for their similarity to the standard
Johnson $B$ and $V$ bandpasses for which there is a substantial heritage
of historical data on Pluto.  The exposure times were chosen to give
comparable signal levels on Charon (a neutrally colored object) but
no attempt was made to adjust exposure times based on the lightcurve
from Pluto.  The signal level expected in the peak pixel on Pluto was
roughly half of the full-well of the detector, leaving ample room to
accommodate Pluto's lightcurve without saturating.  All F435W exposures
were 12 seconds and all F555W exposures were 6 seconds.  Peak counts on
Pluto ranged from 1900 -- 3500 counts and Charon peak counts were from
700 -- 2800 counts.

\placetable{table_circumstances}

A total of 16 images were collected at each visit in each filter,
using a customized dither pattern that provided a 4 by 4 sub-pixel
grid superimposed on a 1.2 arcsec (48-pixel) pattern.  This pattern was
designed to enable the removal of both large and small scale pixellation
effects in the image since the PSF of the telescope is undersampled by the
HRC detector.  The details of the dither pattern are not important for
this project except to note distortion corrections are necessary during
the processing of the data.  If this is not done, the differential
distortion in the dither set will lead to a slight blurring of the
effective PSF in the co-added images.  However, in some visits, the new
satellites can be seen even without removing the differential distortion
and this crude level of stacking was used for the confirmation alluded
to in \citet{wea05}.

Two other important details about the data set should be noted.  First,
since the images span a full year, the data were collected at a range
of solar phase angles, as well as various heliocentric and geocentric
distances.  The signal-to-noise ratio of object images in the data are
clearly inversely proportional to distance and phase angle.  Second,
the data were collected with the largest possible range of roll angle
from visit to visit.  The first point led to an useful variation in
parallax since the system was viewed from a slightly different orientation
during each visit.  The second point helped alleviate potential systematic
effects from the geometric distortion and slightly asymmetric PSF inherent
in the camera.

\section{Analysis}

The new satellites are not directly visible in individual data frames.
Note that there are almost 10 stellar magnitudes difference in brightness
between Pluto and the faintest satellite.  To detect the satellites we
had to co-register and co-add the images from a visit.  When the data
are stacked for each filter separately the objects are visible in many
visits in each filter but the signal-to-noise ratio is low.  To provide
the best possible images for astrometric measurements we chose to co-add
all 32 images (both filters) from each orbit into a single image for
measurement.

As mentioned earlier and shown in Table~\ref{table_circumstances}, the
duration of each visit was somewhat variable.  The shortest visit did
fit in a single visibility window and thus spans only 43 minutes from
start to finish.  The longest visit was spread over three visibility
windows and spanned over 200 minutes.  This time span leads to some
smearing of the satellite images.  The upper limit to this smearing is
about three HRC pixels for the longest observational time span if it
occurred at conjunction.  In practice, the amount of smear is smaller
than this and depends on where in its orbit the satellite appears.

%

The focal plane of ACS not perpendicular to the optical axis, so a
rectangle on the sky looks more like a rhombus in a raw HRC image.
In the interest of preserving the maximum sensitivity on these faint
objects we chose to implement our own method for removing the geometric
distortion from the images by using the forward and inverse distortion
coefficients\footnote{
    More information about the reference files can be found at
    \url{http://www.stsci.edu/hst\slash acs\slash analysis\slash
    reference\_files/idc\_tables.html}.  We used files n7o1634cj\_idc.fits
    and n721640fj\_idc.fits for observations made after 2002 Oct 21 or
    2003 Mar 1, respectively.}
that were available from the STScI website and documented in \citet{gon05}.

To rectify the data, each individual image was resampled onto a
rectilinear grid.  To do this we used the inverse distortion coefficents
to map pixel positions in an orthogonal grid back to CCD pixels in the
HRC's skewed grid.  We defined an undistorted grid with a platescale
of 0.025"/pixel and a sub-sampling factor of 8 (virtual platescale of
0.003125"/pixel).  The flux of each sub-sampled pixel was assigned the
value from the distorted image where the position of the sub-sampled pixel
is mapped.  In this way the entire image on the sub-sampled grid is filled
from values taken from the original image.  Finally, the undistorted,
subsampled images were rebinned to the final output platescale by
averaging the flux in the 8 by 8 grid mapping to each output pixel.

Once the distortion was removed, the position of Charon was measured
using a synthetic photometry aperture of 2.5-pixel radius and no attempt
was made to correct for the variable PSF wing of Pluto at the position
of Charon.  A test fit was performed on the ACS-based astrometry for
Charon to look for effects caused by PSF overlap.  If present, the errors
should show a double-peaked signature when phased by orbital longitude
(maximum error at each minimum separation).  This pattern was not seen
and we conclude that the PSF-overlap errors are negligible compared to
other sources of error in the astrometry.

The images were then stacked by nearest pixel registration based on
Charon's position.  Two sets of stacked images were produced, one
was a straight sum of all frames and the other was a robust average
(sigma-clipping algorithm) meant to suppress cosmic-ray strikes and
other image imperfections.  In the robust average the cores of Pluto and
Charon do not stack properly and do not form useful images.  Therefore,
positions of Pluto and Charon were measured from the straight sum and
positions of the faint satellites were measured from the robust average.

All raw positional measurements were made on undistorted frames
that are rotated by some angle relative to the sky.  We converted all
raw measurements to a J2000 sky-plane measurement by rotating by the
angle given in the ORIENTAT keyword in each image header.  This rotation
angle is held constant within a visit by the tracking procedure employed
by HST\null.  Any error in this angle is neglected though it will be
included in the aggregate error during the fitting process since each
visit has its own independent rotation.  Additionally, all measurements
are treated as relative measurements.  The astrometric zero-point
reference for each image provided in the image headers is not accurate
enough for orbit fitting.  The position of all four objects are accurate
relative to each other within a frame.  Charon was used throughout as a
reference point from which the frame-relative measurements tied together
into a single astrometric system.

We chose not to use Pluto as a registration object due to its larger and
resolved size.  Also, Pluto's substantially larger lightcurve amplitude
relative to Charon \citep{bui97} led us to use Charon to help minimize
errors that might be introduced by center-of-light to center-of-body
offsets.

The positions of Pluto, Charon, P1, and P2 were measured manually in the
stacked images.  The manual measurement was discretized at 1/10 pixel
and done by drawing a 2.5-pixel radius circle on the image.  When the
circle was judged to be in the correct place based on a highly zoomed
image with a logarithmic display stretch, the position was recorded.
In the case of Pluto and Charon, the general location is quite obvious
and the manual measurement can attempt to correct for systematic image
effects (eg., overlapping PSFs).

Measuring P1 and P2 was also done with manual centering but identifying
the region of interest is much more important.  Analogous to moving object
detection for Near-Earth object searches or Kuiper Belt surveys, having
a geometric constraint enables identification of objects at a much lower
signal-to-noise ratio via combining information from multiple epochs.
The probability of a chance coincidence across visits that obeys a
Keplerian orbit vanishes as the number of visits increases.

A crude predictor was used to identify where to look in the images.  The
first step was to draw projected ellipses on the image consistent with the
discovery information (as found in IAUC 8625)\null.  The semi-major axes
were used: $a_{\rm P1}$ = 64700~km and $a_{\rm P2}$ = 49400~km as well
as the assumption of co-planar and circular orbits.  From this guide we
scanned the images for faint objects at any longitude near the projected
ellipses of the orbits.  Visit~7 showed the most convincing images
of objects similar in brightness to that expected for P1 and P2\null.
The reality of the detections was made even more convincing when the
F435W and F555W images were stacked separately and the objects appeared
in both filters.  This detection formed the basis for our confirmation
of the existence of P1 and P2 as reported in IAUC 8625 and \citet{wea05}.

Other images also showed possible detections.  To check if these apparent
detections were real, we considered whether their longitudes were
consistent with possible Keplerian orbits.  Starting from the initial
orbital periods of the discovery report, we adjusted the periods so as
to reproduce the positions of the satellites in Visits 1 and 7, arriving
at 38.25 days for P1 and 24.85 days for P2\null.  From this information
we could crudely predict the locations of the satellites on all frames
(stacked relative to Charon) and highlight a 10$\times$10 pixel region
of interest on those images.  We then located and measured relative to
Charon the position of the most convincing source within that region of
interest.  In all 12 visits a source was identified for P1 and sources
were identified with P2 in 8 visits.  These measurements, relative
to Charon's location, formed the basis for our initial unrestricted
orbital fits.

We also needed the position of Pluto, to estimate the location of the
barycenter about which the satellites orbit.  The centroid positions
for Pluto were not used in the orbit fitting process to avoid using
a center-of-light measurement.  However, the primary purpose of this
data set was to determine an albedo map for the surface of Pluto
\citep{bui05}\null.  This map allows for a more precise determination of
the center of body during the map fitting process.  A comparison of the
two measurements shows a shift of 2 to 15~mas east and $-$10 to 17~mas
in declination.  We used the location of Pluto relative to Charon in
conjunction with the Charon/Pluto mass ratio of 0.122 from \citet{olk03}
for an initial estimate of the location of the barycenter.

Orbits for P1 and P2 about the \citet{olk03} barycenter were fitted
using a downhill simplex minimization scheme \citep{nel65}, using
code developed for fitting orbits of binary trans-neptunian objects
\citep{nol04a, nol04b}.  Initial results from these fits looked very
promising, with residuals mostly at or below a single HRC pixel (0.025
arcsec).  At this step the measurements of P1 from visits 3 and 10 were
excluded from the fit due to excessive residuals, 18 and 31$\sigma$
respectively.

We expected that additional detections would be enabled by eliminating the
differential smear between Charon and the new satellites.  Our initial
fitted orbits were used to predict the locations of P1 and P2 relative
to Charon.  From these positions, offsets were computed enabling us to
stack the images on each satellite in turn, rather than stacking relative
to Charon as was done before.  In the resulting P1 and P2 stacked images,
we identified sources for the satellites in all 12 visits.  This second
generation astrometry went into a second round of unrestricted orbit
fits, again relative to the \citet{olk03} barycenter.  From the residuals
relative to these new orbits it was apparent that one measurement of P2
and two of P1 had unacceptably large residuals and those measurements
were removed from further consideration.  At this point we had precise
astrometric measurements of Pluto, Charon, P1, and P2, relative to one
another.  These data are tabulated in Table~\ref{table_observations}\null.

\placetable{table_observations}

Next we fit completely unrestricted orbits to the data for Charon, P1,
and P2\null.  To extend the time baseline, and thus improve the constraint
on the orbital periods, we included data from two additional sources.
For Charon, 60 positions from \citet{tho97} were combined with our
394 positions (Charon was measurable in each individual frame, unlike
P1 and P2 which could only be measured in the stacked images from each
visit).  For P1 and P2, our positions were augmented with the
two \citet{wea05} positions.  In both cases, the additional data were
collected with different observing strategies and different instruments,
and thus have different potential systematic issues.  For instance,
the \citet{tho97} data were measured relative to the center of light
of Pluto, not the center of body.  Likewise the \citet{wea05} positions
were measured relative to the center of light of Pluto as deduced from
its diffraction spikes, since Pluto itself was severely saturated.

\placetable{table_elements}

Table~\ref{table_elements} summarizes our best fit orbital elements.
Orbital element uncertainties were estimated for each parameter by fixing
the parameter in question at a series of values straddling the best value,
and for each of those values, allowing all other parameters to adjust
themselves to re-minimize $\chi^2$\null.  This process produced a slice
through $\chi^2$ space.  According to \citet{numrec92}, $\chi^2_{\rm
minimum}+1$ is the 1-$\sigma$ confidence contour in this space,
the location of which we report as an uncertainty.  In instances of
asymmetric $\chi^2$ minima, we conservatively report the uncertainty
computed for the shallower-sloped side of the valley.

We performed test fits restricted to just our new ACS astrometry for all
three satellites.  The only significant change seen in the orbit fits were
larger errors on the periods.  However, each orbit fit provides a measurement
of the system mass when combining the period and semi-major axis.  The fit
to the combined data set for Charon gives
$M_{\rm total} = 1.4570 \pm 0.0009$~x~$10^{22}$~kg and is
the value we adopt for the remainder of this work.  The mass inferred from
the P1 orbit is 1.4765~$\pm$~0.006~$10^{22}$~kg and the mass from the P2
orbit is $1.480 \pm 0.011$~x~$10^{22}$~kg.  The P1- and P2-based masses agree
with each other but do not completely agree with the Charon-based mass.
So far we have been unable to explain this discrepancy and its resolution
is left for future work.

\section{Results}


The projected orbits for Pluto, Charon, P1, and P2 are shown in
Figure~\ref{fig_ellipses}\null.  Points with error bars show the sky
plane positions of the observations while the open circles indicate the
locations computed from our fitted orbits.  The degree to which the
circles are centered on the symbols indicates the quality of the fits.
One can also see from this figure the effect of the changing geometry
through the 12 months of observation, and especially for the \citet{wea05}
positions obtained 2 years later.  The data (and the fitted orbit
positions) do not exactly track the instantaneous apparent ellipse.
This figure also shows that we managed to get reasonably complete
longitude coverage of both new satellites.

\placefigure{fig_ellipses}

\citet{wea05} noted that the new satellites P1 and P2 orbit near mean
motion resonances with Charon and with each other.  Our improved orbit
determinations confirm that the orbital periods are indeed near integer
ratio commensurabilities.  However, our uncertainties preclude the
precise ratios for simple resonances.  Determination of resonant motion
will require a full description of the dynamical state of this four-body
system using orbital integration calculations.  These calculations are
beyond the scope of this paper and remain for future work.  Nonetheless,
if these objects do inhabit resonances, the osculating elements must
vary with time to maintain an oscillating resonant angle.
A simple two-body calculation cannot reveal the nature of these
mean-motion resonances nor can it determine the period of oscillations.
We can use a two-body calculation to calculate the time for the resonant
argument to circulate by 2$\pi$ and thus provide a crude upper limit to
the time-scale for the resonant libration.

The orbital period of P1 is 38.2065$\pm$0.0014 days, while six times the
period of Charon is 38.3234~days.  While this is the period ratio most
nearly commensurate, this 0.3\% difference would lead to circulation
of the resonant argument in 2090$\pm$80~days, less than six years.
Likewise, our period of P2 is 24.8562$\pm$0.0013 days, compared with
4 times the period of Charon, which is 25.5489~days, corresponding to
a 2.7\% difference, and thus the resonant argument will circulate in
only 229$\pm$2days.  Comparing the periods of P1 and P2, we find their
ratio is 1.53710$\pm$0.00006, not the exact ratio of 3/2\null.  Again,
circulation would be quite rapid, 515$\pm$6~days.  These timescales for
resonant libration seem short but the mass distribution of this system
is quite unusual.  The dominant mass of Pluto$+$Charon can be viewed
as if it were a highly asymmetric mass that provides a strong periodic
driving force.  The gravitational force exerted by Pluto on either P1 or
P2 varies by roughly 15\% (peak-to-peak).  This periodic driving force
may well control the nature of resonances in the system on an unusually
short time-scale and certainly deserves further scrutiny.


Eccentricities and inclinations of the satellite orbits will offer
important constraints on possible resonances.  The eccentricity of the
orbit of P1 (0.0052$\pm$0.0011) is significantly non-zero, unlike the
orbits of Charon and P2 which are consistent with zero eccentricity.
Figure~\ref{fig_poles} provides our best determination of the orbit
poles, showing the 1-$\sigma$ contours for the pole positions for
the three satellites, as well as for the Charon orbit determined by
\citet{tho97}\null.  The mild discrepancy between the two Charon poles
may be an artifact of the lack of precise center-of-body measurements
for Pluto in the earlier measurements.  The same systematic offset
can explain the apparently special alignment of the line of apsides in
the orbit fitted to the earlier data as well as the apparent non-zero
eccentricity of that orbit.

\placefigure{fig_poles}



We note that every 35.57 days P1, P2, and the barycenter all line
up, alternating between P1 and P2 both being on the same side of the
barycenter versus being on opposite sides of it.  This 35.57 day interval
corresponds to one half of the difference of the mean motion of P1 and P2,
and is does not require resonances among the satellites.  Charon orbits
much faster than P1 and P2, so the Charon-Pluto line sweeps across
the P1-P2-barycenter line within a few days of each of P1-P2-barycenter
alignment, providing opportunities when all four objects lie nearly along
the same line.  Depending on the date of its flyby of the Pluto system,
New Horizons, NASA's first New Frontiers mission bound for Pluto and the
Kuiper Belt \citep{ste02}, might be able to take advantage of one of these
alignments to obtain an especially striking family portrait.  The 35.57
day interval is shorter than the period of P1, so the orientation of
each successive alignment shifts by about 27\degr\ in orbital longitude.



Figure~\ref{fig_residuals} shows details of the residuals from the
orbit fits for the HST data.  The scatter for Charon is quite low in
the fits but slightly better for our new data.  The mean residual is
3.7~mas and maximum residual is 8~mas in our Cycle 11 data.  The data
from \citet{tho97} had a mean residual of 5~mas and a maximum residual of
11~mas.  The lower panels of Figure~\ref{fig_residuals} show the residuals
for P1 and P2 on the same scale.  The scatter is noticeably higher owing
to the much lower signal-to-noise images but is still quite respectable,
averaging about 9~mas for P1 and 17~mas for P2 (about half of a pixel).
The \citet{wea05} residuals (circled) are larger on average, owing to
the lower spatial resolution of those measurements.

\placefigure{fig_residuals}

While fitting for each satellite's orbital elements, we also solved
for the Charon/Pluto mass ratio, which determines the location of the
barycenter.  Allowing the mass ratio to be a free parameter in these
fits resulted in two independent mass ratio estimates, one from the
orbit of P1 and one from the orbit of P2\null.  The resulting $\chi^2$
slices, converted to reduced $\chi^2$ by dividing by the number of
degrees of freedom, are shown in Figure~\ref{fig_chisq}\null.  Our best
determination comes from combining both fits with a resulting mass
ratio of 0.1165$\pm$0.0055, consistent with the \citet{olk03} value
of 0.122$\pm$0.008.  When combined with the new occultation diameter of
Charon from \citet{sic05} of $R=602.5\pm1.0$, we can now determine a much
more accurate density for Charon of 1.66 $\pm$ 0.06 g~cm$^{-3}$ where
the dominant source of error is the mass ratio.  The density of Pluto
is thus 2.03~$\pm$~0.06~g~cm$^{-3}$ assuming a
radius\footnote{Choosing a value for the radius of Pluto is not a simple matter.
  Current measurements from mutual events and stellar occultations do
  not completely agree and explanations of the discrepancies depend on
  the models chosen to interpret the data.  Here we use the number adopted
  in most Pluto map fitting projects with an error bar chosen to include
  the range of model values for the radius from mutual event and stellar
  occultation data.}
of 1153 $\pm$ 10 km.
The radius of Pluto is the dominant source of error and is also the most
poorly understood due to the effects of the atmosphere on occultation
lightcurves.  Even with the relatively poor knowledge of Pluto's radius
it is clear that Charon is significantly less dense than Pluto.

\placefigure{fig_chisq}


Each frame recorded insufficient signal from P1 and P2 to permit
time-resolved photometry, but photometric information could be
obtained by stacking all 192 images of each filter based on the
orbital motion determined previously.  The resulting stacked images
are displayed in Fig.~\ref{fig_stack}\null.  We also stacked images and
extracted globally averaged photometry from Pluto and Charon.  The total
aggregate integration time is 2304 and 1152 seconds for F435W and F555W,
respectively.

\placefigure{fig_stack}


\placetable{table_photometry}

Table~\ref{table_photometry} summarizes the photometric information
extracted from the stacked images.  Values listed without uncertainties
are adopted from \citet{bui97}\null.  ``OBMAG'' is the instrumental
magnitude using the same convention as \citet{sir05}.  The count rates
shown for Pluto are provided as a rough guide to the signal level on
the detector.  Since Pluto is resolved in these data, our small-aperture
fluxes cannot be easily corrected to reliable photometry.  However,
the aggregate PSF for Charon is nearly identical to that for P1 and P2.
``OBMAG'' is the instrumental magnitude derived from the count rates
using the same convention as \citet{sir05}.  This raw photometry was
converted to the $UBVRI$ system using the transformation coefficients of
\citet{sir05}.  However, the aperture corrections required a non-standard
method since the effective PSF was blurred by the stacking process.
To correct for the small aperture we assumed the correction is the same
for Charon, P1 and P2.  The zero-point correction was determined by
using the mean-opposition magnitude for Charon from \citet{bui97} of $V =
17.259$ and $(B-V) = 0.710$\null.  ``OppMag'' refers to the transformed
photometry that is relative to the mean opposition distance for Pluto's
orbit ($r=$39.5 AU, $\Delta = $38.5 AU).  Our new photometric data are
a peculiar mix of information collected at a range of heliocentric and
geocentric distances and solar phase angles.  The signal-to-noise ratio
in the P1 and P2 photometry is not good enough from single visits to
permit extraction of any lightcurve or phase angle behavior.  It is good
enough to permit characterization of globally averaged properties since
the geometries are exactly matched between the three objects.

The lines listed as ``AppMag'' show the photometry corrected to the
circumstances of the discovery of the two satellites as reported in
\citet[their photometry being listed as ``Discovery'']{wea05}\null.
The agreement in the P2 magnitude is excellent while our photometry of
P1 is about 3-sigma fainter than previously reported.  The agreement
(or lack thereof) could easily be affected by lightcurve effects.  If P1
has a large lightcurve we have measured the mean while the discovery
observations could have been near a lightcurve maximum.  A few tenths
of a magnitude are not unreasonable for lightcurves of objects in this
size range.  The $V$ magnitudes for P1 and P2 are almost identical, and
correspond to 22~km radii, if the satellites' $V$ albedos are similar
to Charon's ($\sim$35\%).  The satellites would be larger if their albedos
are lower than Charon's: 4\% albedos correspond to $\sim$65~km radii.

The colors presented here indicate that P1 is a spectrally neutral object
(solar colors within the uncertainties).  This color might indicate a
composition similar to Charon, although Charon is marginally redder.
The color for P2 is slightly redder than Pluto.  Both satellites exhibit
colors which are common in the Kuiper Belt \citep[e.g.,][]{pei04}.  The
pattern with radial distance in the system is a bit harder to understand.
If the colors of P1 and P2 were exogenic and derived from material lost
from either Pluto or Charon it is hard to imagine P2 (the next one out
from Charon) to be colored from Pluto material and not be expressed on
Charon's surface.  Likewise, P1 and P2 are not that far apart yet they
have different colors.  If these objects are collisional fragments of
the Pluto-Charon binary formation process \citep{ste05}, the striking
color difference may be hard to explain.  These new observations also
make it clear that environment alone cannot explain the color diversity
amoung KBOs.  Clearly, this is a profound mystery yet to be resolved.

\section{Acknowledgements}

This work was supported by grant HST-GO-09391.01 from the Space Telescope
Science Institute.  These results would not have been possible without
the expert assistance of the late Andy Lubenow (STScI).  We also thank
the free and open source software communities for empowering us with
many of the tools used to complete this project, notably Linux, the
GNU tools, \LaTeX, and FVWM.


\clearpage

{
\begin{deluxetable}{cccccc}
\tablecolumns{6}
\tablecaption{Circumstances of Observations\label{table_circumstances}}
\tablewidth{0pt}
\tablehead{
  \colhead{Midtime}& \colhead{Visit} &
     \colhead{$r$}& \colhead{$\Delta$}& \colhead{$\alpha$}& \colhead{$\Delta t$}\\
  \colhead{(JD)}& \colhead{ID}& \colhead{(AU)}&
     \colhead{(AU)}& \colhead{(degrees)}& \colhead{(hours)}
}
\startdata
2452436.846680&  1& 30.518& 29.521& 0.36& 1.71\\
2452440.051972&  7& 30.520& 29.527& 0.41& 1.71\\
2452444.261280&  3& 30.521& 29.539& 0.50& 1.74\\
2452458.083533&  5& 30.526& 29.615& 0.86& 1.65\\
2452472.979577&  9& 30.532& 29.751& 1.24& 1.66\\
2452550.688720& 11& 30.561& 30.956& 1.71& 1.65\\
2452688.548688&  6& 30.613& 30.944& 1.73& 0.70\\
2452750.256191&  2& 30.637& 29.983& 1.44& 1.50\\
2452772.616060&  8& 30.646& 29.757& 0.91& 1.49\\
2452787.606155& 12& 30.651& 29.675& 0.51& 0.70\\
2452789.659917&  4& 30.652& 29.668& 0.46& 1.71\\
2452799.257479& 10& 30.656& 29.655& 0.32& 1.53\\
\enddata
\tablecomments{$r$ and $\Delta$ are the heliocentric and geocentric
distance to the Pluto-Charon barycenter.  $\alpha$ is the Sun-barycenter-Earth
(phase) angle.  The visit ID is a number from the original observation
sequence that is in order of increasing Pluto sub-Earth longitude and the
time span from first to last image in a visit is listed under $\Delta t$.
}
\end{deluxetable}
}

\clearpage

{
\begin{deluxetable}{rcrrrrrr}
\tablecolumns{8}
\tablecaption{Differential Astrometry\label{table_observations}}
\tablewidth{0pt}
\tablehead{
  \colhead{Midtime} & \colhead{Visit}& \multicolumn{2}{c}{Charon} &
     \multicolumn{2}{c}{S/2005 P1} & \multicolumn{2}{c}{S/2005 P2}\\
  \colhead{(JD)}& \colhead{ID}&
     \colhead{$\Delta\alpha$}& \colhead{$\Delta\delta$}&
     \colhead{$\Delta\alpha$}& \colhead{$\Delta\delta$}&
     \colhead{$\Delta\alpha$}& \colhead{$\Delta\delta$}
}
\startdata
2452436.846680&  1& -0.4410&  0.1192&  1.5552&  0.3647&  0.9313&  1.8835\\
2452440.051972&  7&  0.4501& -0.1106&  1.6298&  1.7786&  0.1570&  2.1420\\
2452444.261280&  3& -0.4292& -0.6855&{\it 0.7003}&{\it 2.8137}& -1.0598& 0.4708\\
2452458.083533&  5& -0.0046& -0.8393& -1.6193& -1.4142&  1.1989&  0.1416\\
2452472.979577&  9&  0.4322&  0.6662&  1.3802& -0.5628& -1.0678& -1.3731\\
2452550.688720& 11&  0.0244&  0.8020&  1.4565&  0.0808&{\it -0.4552}&{\it -2.0719}\\
2452688.548688&  6&  0.2570& -0.5580& -1.6226& -1.7965&  0.1714&  1.9567\\
2452750.256191&  2& -0.5371& -0.2559&  0.8133&  2.8203& -0.3261& -2.1782\\
2452772.616060&  8&  0.5422&  0.2580&  0.0662& -2.6896& -0.9587& -1.9108\\
2452787.606155& 12& -0.2843&  0.5381&  1.0365&  2.8517&  0.2195&  2.2236\\
2452789.659917&  4& -0.2797& -0.8646&  0.5195&  2.8330& -0.4735&  1.6385\\
2452799.257479& 10&  0.2771&  0.8657&{\it -1.6553}&{\it -0.0007}& -0.5006& -2.1083\\
\enddata
\tablecomments{The times for all measurements are the mean of the exposure
mid-times for all combined images.  All offsets are arcseconds in J2000
coordinates relative to the center of Pluto.  Values in italics were not
used in our orbit fits because of anomalously high residuals.
}
\end{deluxetable}
}

\clearpage

{
\begin{deluxetable}{lccc}
\footnotesize
\tablecaption{Orbital elements from unrestricted fits (epoch = 2452600.5)
\label{table_elements}}
\tablewidth{0pt}
\tablehead{
  \colhead{} & Charon & S/2005 P2 & S/2005 P1
}
\startdata
Period (days)		& 6.3872304(11)	& 24.8562(13)	& 38.2065(14) \\
Semi-major axis, $a$ (km)%
			& 19,571.4(4.0)	& 48,675(121)	& 64,780(88) \\
Eccentricity, $e$	& 0.000000(70)	& 0.0023(21)	& 0.0052(11) \\
Inclination, $i$ (deg)	& 96.145(14)	& 96.18(22)	& 96.36(12) \\
Lon.\ ascending node, $\Omega$ (deg)%
			& 223.046(14)	& 223.14(23)	& 223.173(86) \\
Lon.\ periapsis, $\tilde{\omega}$ (deg)%
			&	---	& 216(13)	& 200.1(3.7) \\
Mean lon.\ at epoch, $l$ (deg)%
			& 257.946(13)	& 123.14(20)	& 322.71(23) \\
\enddata
\end{deluxetable}
}

\clearpage

{
\begin{deluxetable}{lccccc}
\footnotesize
\tablecaption{Photometry \label{table_photometry}}
\tablewidth{0pt}
\tablehead{
  \colhead{}&
  \colhead{}&
  \colhead{Pluto}& \colhead{Charon}& \colhead{S/2005 P1}& \colhead{S/2005 P2}
}
\startdata
photons/sec& F435W& 4748.1 $\pm$ 1.5&      1095 $\pm$ 0.7&  1.641 $\pm$ 0.029&  1.195 $\pm$ 0.025\\
           & F555W& 11136.3 $\pm$ 3.2&     2101.8 $\pm$ 1.4& 2.956 $\pm$ 0.055& 2.765 $\pm$ 0.053\\
OBMAG&       F435W&              ---& -7.599 $\pm$ 0.001& -0.538 $\pm$ 0.019& -0.193 $\pm$ 0.023\\
     &       F555W&              ---& -8.306 $\pm$ 0.001& -1.177 $\pm$ 0.020& -1.104 $\pm$ 0.021\\
OppMag&      $B_m$&              ---& 17.259&             25.036 $\pm$ 0.019& 25.357 $\pm$ 0.023\\
      &      $V_m$&              ---& 17.969&             24.393 $\pm$ 0.020& 24.546 $\pm$ 0.021\\
AppMag&        $B$&              ---& 16.903&             23.970 $\pm$ 0.019& 24.291 $\pm$ 0.023\\
      &        $V$&              ---& 16.193&             23.327 $\pm$ 0.020& 23.384 $\pm$ 0.021\\
      &    $(B-V)$&            0.868&  0.710&              0.644 $\pm$ 0.028&  0.907 $\pm$ 0.031\\
Discovery&     $V$&              ---&    ---&             22.93  $\pm$ 0.12&  23.38 $\pm$ 0.17\\
\enddata
\tablecomments{Uncertainties are based on photon-counting statistics
for the object and measured noise in the sky background.  Scaling from
instrumental to absolute magnitudes is done using the known mean magnitude
for Charon from \citet{bui97}.  Magnitudes for Pluto are not reported
due to indeterminant aperture corrections for a resolved object.}
\end{deluxetable}
}

\clearpage
\centerline{\bf FIGURE CAPTIONS}

\figcaption[]{\label{fig_ellipses}\capellipses}

\figcaption[]{\label{fig_poles}\cappoles}

\figcaption[]{\label{fig_residuals}\capresiduals}

\figcaption[]{\label{fig_chisq}\capchisq}

\figcaption[]{\label{fig_stack}\capstack}


\begin{figure}
\figurenum{\ref{fig_ellipses}}
\vskip -0.3truein
\epsscale{1.6}
\hbox{\hskip .2truein\plotone{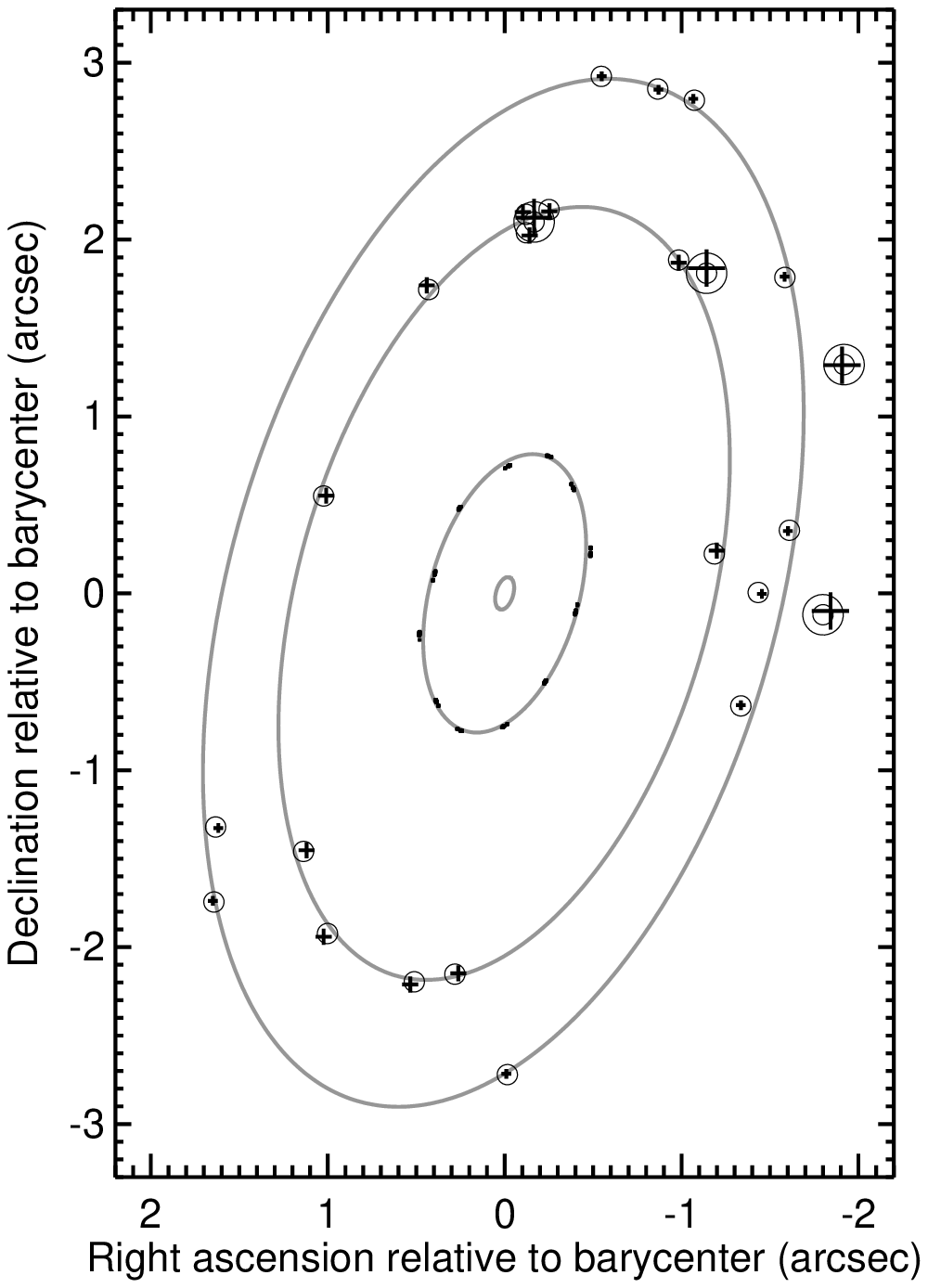}}
\vskip -0.2truein
\caption{\capellipses}
\end{figure}

\begin{figure}
\figurenum{\ref{fig_poles}}
\vskip -0.8truein
\epsscale{1.29}
\hbox{\hskip -.3truein\plotone{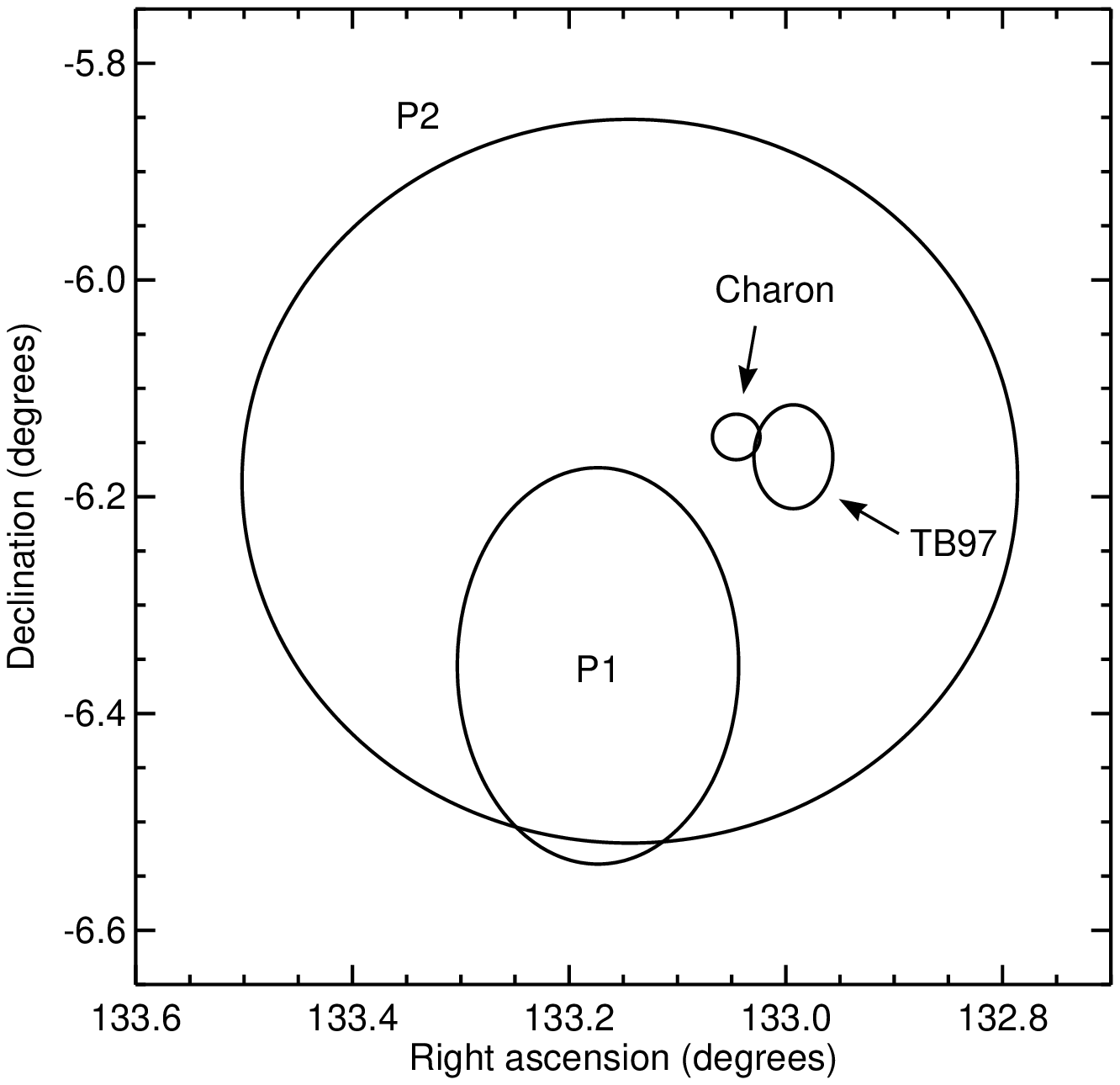}}
\vskip -0.2truein
\caption{\cappoles}
\end{figure}

\begin{figure}
\figurenum{\ref{fig_residuals}}
\vskip -0.2truein
\epsscale{0.85}
\hbox{\hskip 0.20truein\plotone{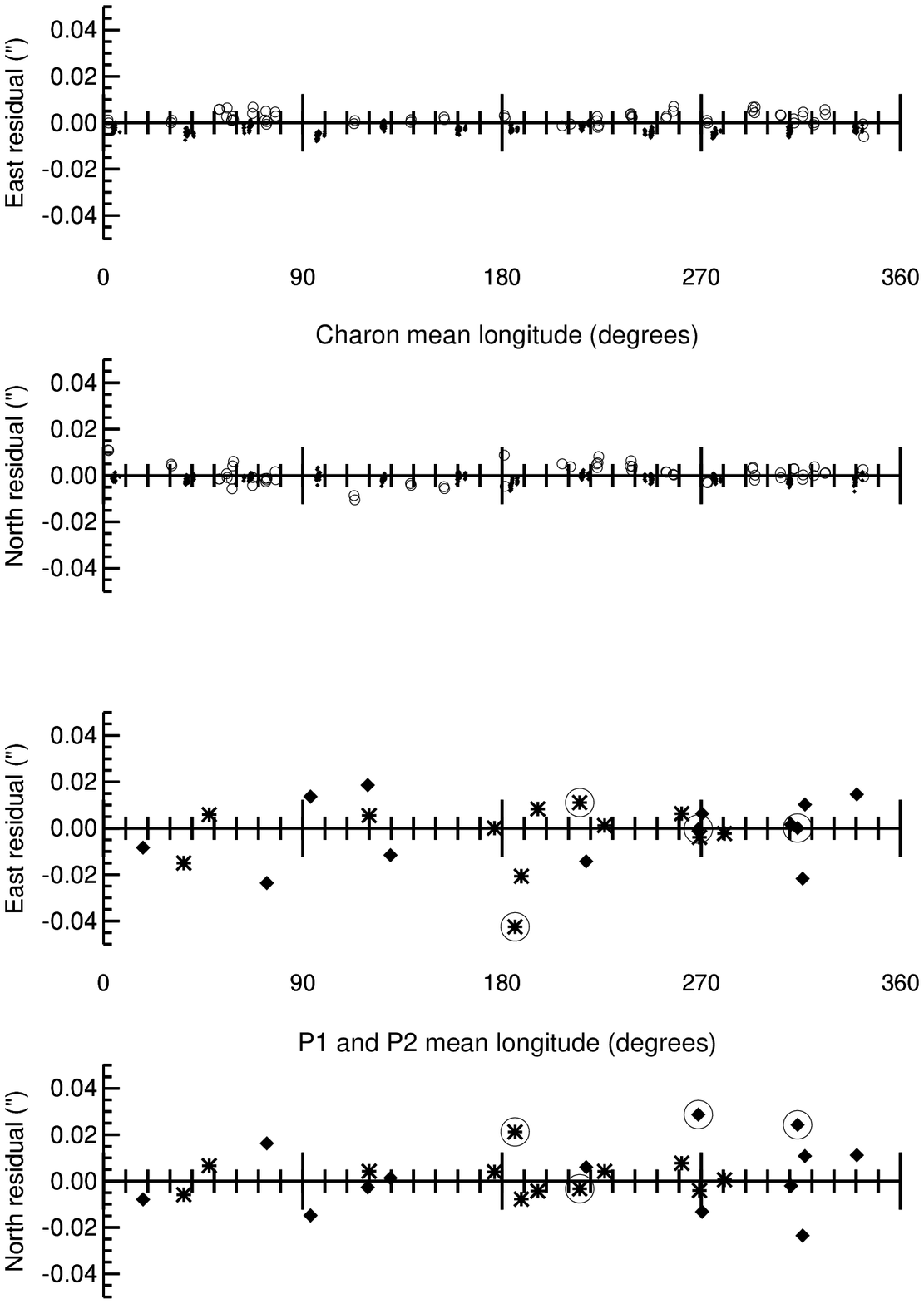}}
\vskip -0.2truein
\caption{\capresiduals}
\end{figure}

\begin{figure}
\figurenum{\ref{fig_chisq}}
\vskip -0.8truein
\epsscale{1.05}
\hbox{\hskip -.3truein\plotone{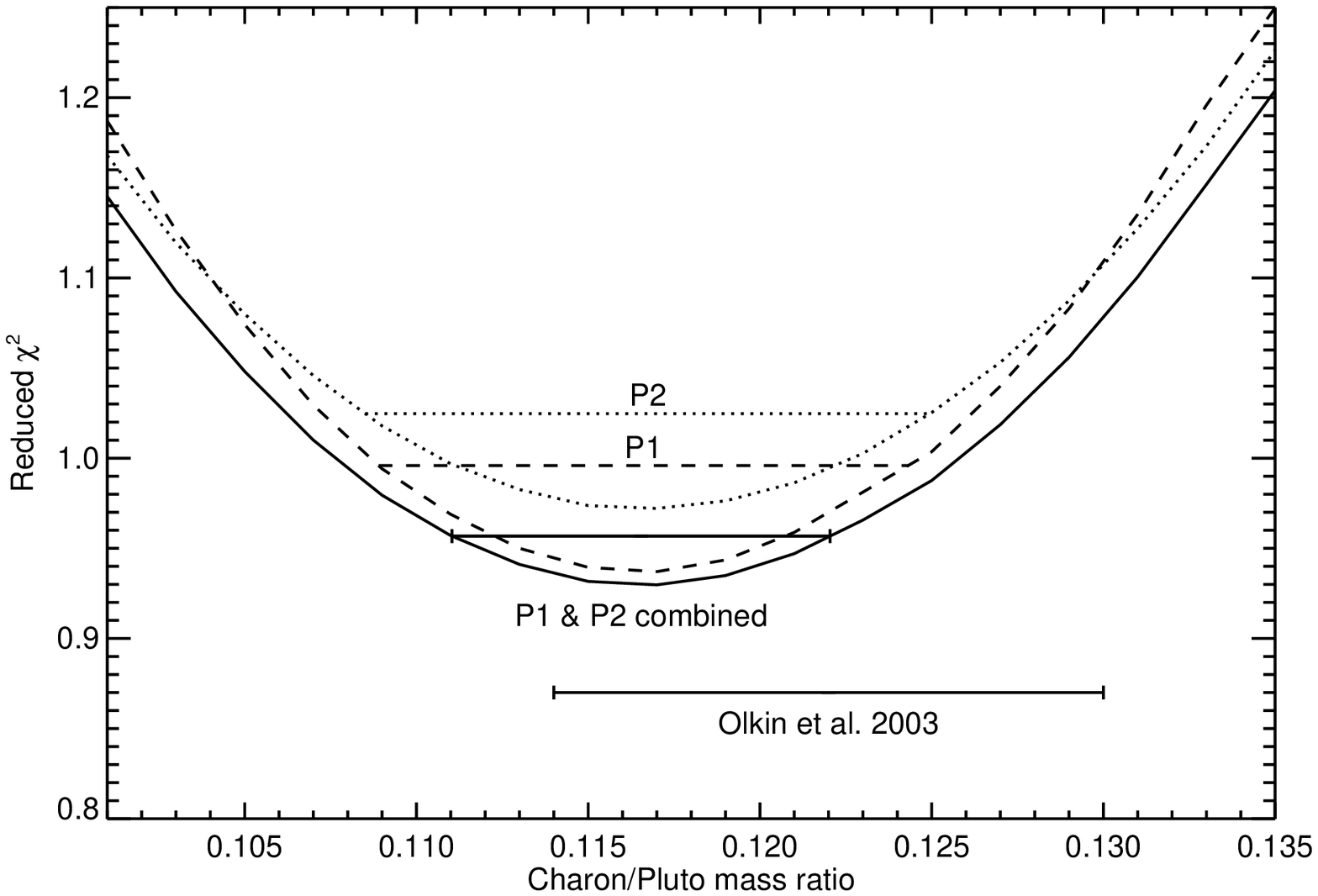}}
\vskip -0.2truein
\caption{\capchisq}
\end{figure}

\begin{figure}
\figurenum{\ref{fig_stack}}
\vskip -0.8truein
\epsscale{1.05}
\hbox{\hskip -.3truein\plotone{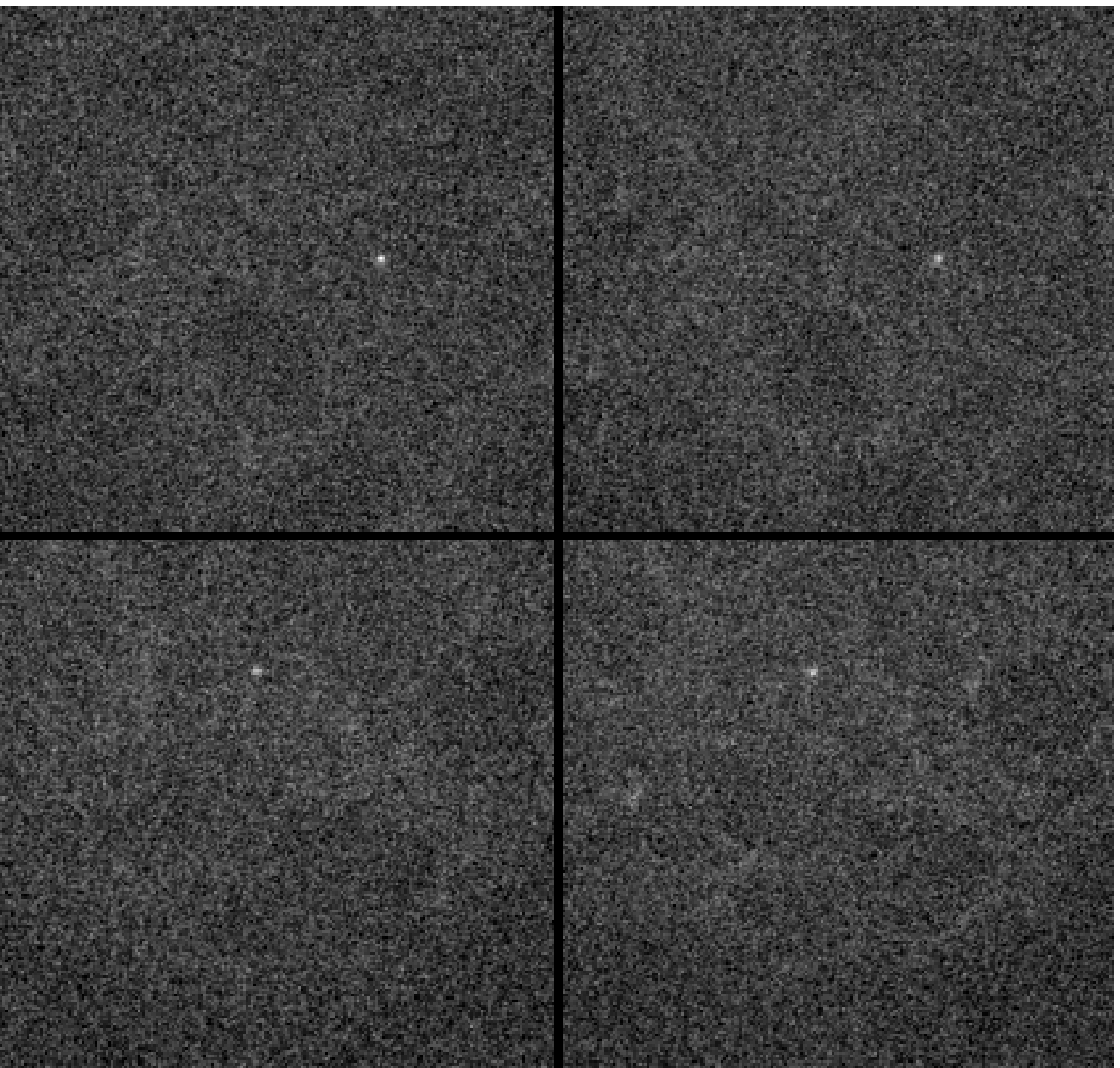}}
\vskip -0.2truein
\caption{\capstack}
\end{figure}

\end{document}